\documentclass{cjaa}                   
\usepackage{graphicx}                   
\input{epsf.sty}                        
\input{psfig.sty}                       

\begin{document}

   \title{Pulsars and quark stars
}

   \volnopage{Vol.0 (200x) No.0, 000--000}      
   \setcounter{page}{1}          

   \author{Ren-Xin Xu
      }
   \offprints{Ren-Xin Xu}                   

   \institute{Astronomy Department, School of Physics, Peking University, Beijing 100871, China\\
             \email{r.x.xu@pku.edu.cn}
\vspace{5mm}
{\em ~~~~~~~~~~Not only is the Universe stranger than we imagine,
it is stranger than we {\bf can} imagine.} \\
~~~~~~~~~~~~~~~~~~~~~~~~~~~~~~~~~~~~~~~~~~~~~~~~~~~~~~~~~~~~~~~~
{------~~Sir Arthur Stanley Eddington ($1882\sim 1944$)}
          }

\date{Received~~~~~~~~~~~~~~~~~; ~~accepted}

   \abstract{
Members of the family of pulsar-like stars are distinguished by
their different manifestations observed, i.e., radio pulsars,
accretion-driven X-ray pulsars, X-ray bursts, anomalous X-ray
pulsars/soft gamma-ray repeaters, compact center objects, and dim
thermal neutron stars.
Though one may conventionally think that these stars are normal
neutron stars, it is still an open issue whether they are actually
neutron stars or quark stars, as no convincing work, either
theoretical from first principles or observational, has confirmed
Baade-Zwicky's original idea that supernovae produce neutron
stars.
After introducing briefly the history of pulsars and quark stars,
the author summarizes the recent achievements in his pulsar group,
including quark matter phenomenology at low temperature,
starquakes of solid pulsars, low-mass quark stars, and the pulsar
magnetospheric activities.
   }

   \authorrunning{Xu}            
   \titlerunning{Pulsars and quark stars}  

   \maketitle
%
\section{Introduction}

Astronomers uncovered compact white dwarfs as early as in 1914,
but were not clear how these objects support themselves against
gravity. It is Ralph Howard Fowler ($1889\sim 1944$) who
recognized the quantum pressure of degenerate electrons in white
dwarfs, only about two months later after Dirac's paper on the
Fermi-Dirac distribution in 1926.
Chandrasekhar obtained the equation of state of completely
degenerate electron gas, with the inclusion of special
relativistic effect, calculated hydrostatic equilibrium of stars
composed of this matter, and then found that the degenerate
pressure is not omnipotent in standing against the gravitational
collapse (i.e., there exists a mass-limit beyond which a white
dwarf can not be hydrostatic) in 1931.
What if the mass of a star supported by electron degenerate
pressure is greater than the Chandrasekhar limit?
Landau predicted a state of matter, the density of which ``becomes
so great that atomic nuclei come in close contact, forming one
{\em gigantic nucleus}'' in 1932. A star composed dominantly of
such matter is called a ``neutron'' star, and Baade and Zwicky
suggested in 1934 that supernovae could produce neutron stars. A
direct observational evidence, proposed by Gold in 1968, is
detecting pulsed radio beams ({\em pulsars}) due to the lighthouse
effect of spinning neutron stars, although pulsars were supposed
to ``be associated with oscillation of white dwarfs or neutron
stars'' when discovered by Hewish, Bell, and their co-authors.

Are neutrons elementary particles?
A success in the classification of hadrons discovered in cosmic
rays and in accelerators leaded Gell-Mann (1964) to coin ``{\em
quark}'' with fraction charges ($\pm 1/3, \mp 2/3$) in
mathematical description, rather than in reality.
These sub-nucleon particles were treated independently as real
components of hadrons by Zweig (1964, called as ``{\em ace}'') and
by the Chinese group (1966, called as ``{\em straton}''). It is
suggested that quarks have colour-charges (an analogy that an
electron has electricity-charge), and that meson and baryon are
bond states of quark-antiquark and of 3-quark, respectively. All
the six flavors of quarks ($u,d,c,s,t,b$) have experimental
evidence (the evidence for the last one, top quark, was reported
in 1995).
What's kind of interaction between quarks? The underlying theory
is believed to be quantum chromodynamics (QCD), a non-Abelian
$SU(3)$ gauge theory. Gross \& Wilczek (1973) and Politzer (1973)
noted in QCD that the effective coupling between quarks decreases
with energy (the {\em asymptotic freedom}), which was found to
agree with the SLAC (the linear accelerator at Stanford)
experiments in 1960s.
Quark matter (or quark-gluon plasma), the soup of deconfined
quarks and gluons, is a direct consequence of asymptotic freedom
when temperature or baryon density are high enough since quarks
are more fundamental than neutrons (or protons).

Are pulsars really neutron stars?
Quarks in nucleons (neutrons$=\{udd\}$ and protons$=\{uud\}$) may
deconfine to become quark matter at supranuclear density because
of asymptotic freedom, but a real question is at what density the
deconfinement occurs.
Due to a mathematical complex of the nonlinear nature of QCD at
low energy, one can {\em not} tell us the exact density of
phase-transition between hadron and quark-matter phases from first
principles.
Nevertheless, the density could be only $\varrho_{\rm c}\simeq
(4\pi r_{\rm n}^3/3)^{-1}\simeq 1.5\varrho_0$ ($\varrho_0=0.16$
fm$^{-3}$ is the nuclear density) {\em if} nucleon keeps a radius
$r_{\rm n}\sim 1$ fm since the vacuum outside nucleons decreases
as baryon number density gets higher and higher.
The center density in most normal neutron star models can reach
$\varrho_{\rm c}$, and quark-matter cores in neutron stars began
to be proposed in 1969, long before knowing asymptotic freedom.
If quark matter can be stable at zero pressure, {\em quark stars},
composed completely by quark matter, may then exist in the
Universe. This possibility become more likely if bulk strange
quark matter with almost equal numbers of $u$, $d$, and $s$ quarks
is absolutely stable (Bodmer 1971, Witten 1984).
As strange (quark matter) stars can easily reproduce the rotation
and emission features of radio pulsars, it is {\em not} necessary
for us to believe that pulsars are normal neutron stars since
pulsars could alternatively be quark stars.

This paper is a continuation of two previous reviews (Xu 2003a,
2003b), in which recent achievements of quark star as the nature
of pulsar-like compact objects are summarized.

\section{Quark matter phenomenology at low temperature}

Asymptotic freedom results in two distinguished phases in the QCD
phase-diagram (temperature $T$ v.s. baryon chemical potential
$\mu_{\rm B}$): hadron gas and quark matter (separated by
``deconfinement'' in Fig. 1 of Xu 2005a).
However, in different locations of the diagram, the vacuum would
have different features and is thus classified into two types: the
perturbative-QCD (pQCD) vacuum and nonperturbative-QCD (QCD)
vacuum. The coupling is weak in the former, but is strong in the
later.
Quark-antiquark (and gluons) condensations occur in QCD vacuum
(i.e., the expected value of $\langle {\bar q}q\rangle \neq 0$),
but not in pQCD vacuum.
The chiral symmetry is spontaneously broken in case the vacuum is
changed from pQCD to QCD vacuums, and ({\em bare}) quarks become
then massive constituent ones ({\em dressed} quarks).
There is no observation that the quark deconfinement and the
chiral symmetry restoration should occur simultaneously.

Actually two kinds of quark matter are focused in recent studies:
{\em temperature}-dominated ($T\gg 0$ but $\mu_{\rm B}\sim 0$) and
{\em density}-dominated ($T\sim 0$, $\mu_{\rm B}\gg 0$).
Previously, Monte Carlo simulations of lattice QCD (LQCD) were
only applicable for cases with $\mu_{\rm B}=0$. Only recent
attempts are tried at $\mu_{\rm B}\neq 0$ (quark stars or nuggets)
in LQCD. We have then to rely on phenomenological models to
speculate on the properties of density-dominated quark matter.
Empirically, one may think that ordinary matter at low $T$ should
be solidified. Should quark matter be in a solid state at
extremely low temperature ($T\ll 1$ MeV) too?

In the region where quarks are deconfined while the chiral
symmetry is broken, the coupling between dressed quarks is very
strong, which may favor the formation of $n-$quark ($n$: the
number of quarks in a cluster) clusters (Xu 2003c). Such quark
clusters could be very likely in an analogy of $\alpha$-clusters
moving in nuclei, which are well known in nuclear physics.
The quark clusters in quark matter would be {\em classical}
(rather than quantum) ``composite particles'' if the cluster's
wavepackages do not overlap.
Consequently, fluid quark matter would be phase-converted to a
solid one if the thermal energy is much smaller than the
interaction energy between quark clusters. Quark stars at low
temperature could then be conjectured to be solid.

Ferro-magnetization may occur in a solid quark matter, without
field decay in the stars.
Magnetic field plays a key role in pulsar life, but there is still
no consensus on its physical origin although some ideas relevant
(e.g., the flux conservation during collapse, the dynamo action)
appeared in the literatures.
Quark clusters with magnetic momentum may exist in solid quark
stars. Solid magnetic quark matter might then magnetize itself
spontaneously at sufficient low temperature (below its Curie
critical temperature, $T_{\rm curie}$) by, e.g., the
flux-conserved field. Ferromagnetism saturated may result in a
very strong {\em dipole} magnetic field.
We therefore speculate simply a ferromagnetic origin of pulsar
strong magnetic fields. Magnetic fields forming in this way could
not decay since the energy-scale ($T_{\rm curie}$) is much higher
than that of any electric currents inside quark stars or in their
magnetosphere.

\section{Glitches as starquakes of solid pulsars}

A solid stellar object (the most well-studied one is the Earth)
would inevitably result in starquakes when strain energy develops
to a critical value.
It is worth noting that huge energy should be released (and thus
large spin-change occurs) after a quake of solid quark stars
because of the almost homogenous distribution of total stellar
matter with supranuclear density.
Starquakes could surely then be a simple and intuitional mechanism
for pulsar glitches.

Actually, a quake model for a star to be mostly solid was
generally discussed by Baym \& Pines (1971), who parameterized the
dynamics for solid crusts, and possible solid cores, of neutron
stars.
Strain energy develops when a solid star spins down until a quake
occurs when stellar stresses reach a critical value.
As for the solid quark stars, we suggest that, during a quake, the
entire stress is almost relieved at first when the quake cracks
the star in pieces of small size (the total released energy
$E_{\rm t}$ may be converted into thermal energy $E_{\rm therm}$
and kinematic energy $E_{\rm k}$ of plastic flow, $E_{\rm
t}=E_{\rm therm}+E_{\rm k})$, but the part of $E_{\rm k}$ might be
re-stored by stress due to the anelastic flow (i.e., the kinetic
energy is converted to strain energy again).
A quark star may solidify with an initial oblateness (or called
ellipticity) $\varepsilon_0$; stress increases as the star losses
its rotation energy, until the star reaches an oblateness
$\varepsilon_{+1}$ when a quake occurs. The reference point of
strain energy is suggested to be $\varepsilon_1$ (the oblateness
of a star without shear energy) after the glitch, but the real
oblateness could be $\varepsilon_{-1}<\varepsilon_1$ (Zhou et al.
2004).

The density of quark stars with mass $< \sim 1.5 M_\odot$ can be
well approximated to be uniform.
As a star, with an initial value $\varepsilon_{0}$, slows down,
the expected $\varepsilon$ decreases with increasing period.
However, the rigidity of the solid star causes it to remain more
oblate than it would be had it no resistance to shear. The strain
energy and the mean stress $\sigma$ are then (Baym \& Pines 1971)
\begin{equation}
E_{\rm strain}= A_2(\varepsilon  - \varepsilon_{0} )^2,~~~
\sigma  = \left| {\frac{1}{{V_{} }}\frac{{\partial E_{\rm strain}
}}{{\partial \varepsilon }}} \right| = \mu (\varepsilon _0  -
\varepsilon ),%
\label{sigma}
\end{equation}
respectively, where $\varepsilon$ is the real oblateness of a
star, $V=4\pi R^3/3$ is the volume of the star, and $\mu  = 2B/V $
is the mean shear modulus of the star.
The stress could be developed by stellar spindown and by other
factors (see discussions below).

The total energy of a star with mass $M$ and radius $R$ is mostly
the gravitational energy $E_{\rm gravi}$, the rotation energy $ E_
{\rm rot}$, and the strain energy $E_{\rm strain}$,
\begin{equation}
E = E_{\rm gravi} + E_ {\rm rot} + E_{\rm strain}= E_{\rm 0} +
A_1\varepsilon ^2 + L^2/(2I) + A_2(\varepsilon  - \varepsilon_i)^2
\label{E}
\end{equation}
where $\varepsilon_i$ is the reference oblateness before the
$(i+1)$-th glitch occurs, $E_{\rm 0}=-3M^2G/(5R)$, $I$ is the
moment of inertia, $L=I\Omega$ is the stellar angular momentum,
$\Omega=2\pi/P$ ($P$ the rotation period), and the coefficients
$A_1$ and $A_2$ measure the gravitational and strain energies
(Baym \& Pines 1971), respectively,
\begin{equation}
A_1 = \frac{3}{{25}}\frac{{GM_{}^2 }}{R},~~~~~
A_2 = {2\over 3}\pi R^3 \mu.
\end{equation}
By minimizing $E$, a real state satisfies (note that $\partial
I(\varepsilon)/\partial \varepsilon=I_0$),
\begin{equation}
 \varepsilon  = \frac{{I_0 \Omega ^2}}{{4(A_1 +
A_2)}} + \frac{A_2}{{A_1 + A_2}}\varepsilon _i. %
\label{epsilon}
\end{equation}
The reference oblateness is assumed, by setting $B=0$ in
Eq.(\ref{epsilon}), to be
\begin{equation}
\varepsilon _i  = I_0 \Omega^2/(4A_1).%
\label{epsiloni}
\end{equation}
A star with oblateness of Eq.(\ref{epsiloni}) is actually a
Maclaurin sphere. When the star spins down to $\Omega$, the stress
develops to,
\begin{equation}
\sigma = \mu [\frac{A_1}{{A_1 + A_2}}\varepsilon_i - \frac{{I_0
\Omega^2
}}{{4(A_1 + A_2)}} ],%
\label{sigmamu}
\end{equation}
from Eqs. (\ref{sigma}), (\ref{epsilon}), and (\ref{epsiloni}). A
glitch occurs when $\sigma>\sigma_{\rm c}$ ($\sigma_{\rm c}$: the
critical stress).

A detail model in this scenario was introduced in Zhou el al.
(2004), where it is found that the general glitch natures (i.e.,
the glitch amplitudes and the time intervals) could be reproduced
if solid quark matter, with high baryon density but low
temperature, has properties of shear modulus $\mu=10^{30\sim 34}$
erg/cm$^3$ and critical stress $\sigma_{\rm c}=10^{18\sim 24}$
erg/cm$^3$.

{\em Why quake?}
The key point that a solid star differs from a fluid one is stress
energy developed only available in the solid star. Factually, an
elastic body can have two kind of changes: {\em shearing} and {\em
bulk} strains. The volume, $V$, changes in the later, but not in
the former. If both strains are included, Eq.(\ref{E}) would
become,
\begin{equation}
E=E_0+A_1 \varepsilon^2+L^2/(2I)+A_2(\varepsilon  -
\varepsilon_i)^2+K(\Delta V/V_0)^2,
\label{EE}
\end{equation}
where $K$, which is order of $\mu$, is the bulk modulus, $\Delta
V=V-V_0$, and $V_0$ the volume of the body without stress.
Besides the shear strain of ellipsoid change discussed above,
azimuthal stress due to the general relativistic effect (being
similar to the frame-dragging effect in vacuum) of rotating solid
stars may also contribute significantly, though, unfortunately,
the theoretical answers to elastic relativistic-stars with
rotation are very difficult to be worked out.

Elastic energy develops as a solid bare strange star cools
($\Rightarrow$ bulk strain) and spins down ($\Rightarrow$ shearing
strain; even spinning constantly).
The temperature-dependent quantity $\Delta V/V_0\sim (\Delta
R/R_0)^3$, with $R$ the radius, $V_0\simeq 4\pi R_0^3/3$, $T$ the
temperature. The value of $R(T=40{\rm MeV})-R(T=0)$ could be order
of a few hundreds of meters (Blaschke et al. 2004). The giant
frequency glitch in KS 1947+300 could be evidence for a quake
caused by bulk-stress energy release (i.e., bulk elastic force
increases to a critical point), but one may expect a sudden
decrease in the pulse frequency when a star is spinning up in the
glitch model of normal neutron stars (Galloway et al. 2004).
The glitch in KS 1947+300 can be reproduced as long as the star
shrinks with $\Delta R/R=-0.5 \Delta\nu/\nu\sim 10^{-5}$.

{\em Anomalous X-ray pulsars (AXPs) \& Soft $\gamma$-ray repeaters
(SGRs)}.
AXP/SGRs are supposed to be magnetars, a kind of neutron stars
with surface fields of order of $10^{13\sim 14}$ G, or even
higher. But an alternative suggestion is that they are
normal-field pulsar-like stars which are in an accretion propeller
phase. The very difficulty in the later view point is to reproduce
the irregular bursts, even with peak luminosity $\sim 10^7L_{\rm
Edd}$ (SGR 0526-66; $L_{\rm Edd}$ the Eddington luminosity).
Though it is possible that giant bursts may be the results of the
bombardments of comet-like objects (e.g., strange planets) to bare
strange stars, moderate bursts could be of quake-induced.
Both shear and bulk strain-induced quakes could occur in AXP/SGRs
when the stress energy increase to a critical value. Stress energy
as well as magnetic energy (and probably gravitational energy)
could be released during quakes.
The glitches in AXP/SGRs (e.g., 1E 2259+586, around the 2002
outburst) could be examples of such quakes (Kaspi 2004).
For $\sigma_c\sim 10^{22}$ erg/cm$^3$, the total elastic energy
released could be order of $\sim 10^{40}$ erg/cm$^3$ when a quake
occurs in a solid quark star with radius $\sim 10^6$ cm. This
energy is comparable to the observed values for two X-ray flares
in 1E 1048.1-5937 (Gavriil \& Kaspi 2004).

It is suggested (Lyne 2004, Lin \& Zhang 2004) that pulsar-like
stars could evolve from normal radio pulsars to magnetars
(AXP/SGRs) through un-recoverable glitches (neither the period nor
the period derivative are completely recovered for Crab, Vela, and
other young pulsars).
However, the ``appearance'' magnetic field, $B$, increasing could
arise from the shrinking of pulsars after quakes. If the effect of
decreasing $I$ is included, the appearance value of $B$ should be
derived by solving
\begin{equation}
I{\dot \Omega} + {1\over 2}{\dot I}\Omega= -{2\over 3c^3}B^2R^6\Omega^3,%
\label{dotO}
\end{equation}
though the real magnetic filed keeps constantly.
Significant gravitational wave could radiate during a glitch,
which could be detectable in order to test the quake model
presented.

\section{Low-mass quark stars?}

It is well known that the mass of quark stars could range from
$>10^{2\sim 3}$ baryons (strangelets) to $\sim M_\odot$, but the
minimum mass of normal neutron stars is $\sim 0.1 M_\odot$.
For these stars with masses $<M_\odot$, the mass($M$)-radius($R$)
relations are in striking contrast: $M\propto R^3$ for quark stars
due to the self color confinement of quark matter but $M\propto
R^{-3}$ for normal neutron stars due to the gravitational binding.
Therefore, low-mass quark stars would be a direct consequence of
the possible existence of astrophysical quark matter.

Up-to-now, there are 7 kinds of members in the family of
pulsar-like stars discovered: radio pulsars (normal pulsars and
millisecond pulsars), accretion-powered X-ray pulsars, X-ray
bursts, AXPs, SGRs, compact center objects (CCOs), and dim thermal
``Neutron'' stars (DTNs).
Could there be any hints that some of these compact objects are
actually low-mass quark stars?
It is admitted that all the determined masses in binaries with
pulsars are $\sim 1.4M_\odot$. Nonetheless, observationally, the
existence of low-mass pulsar-like stars may still not be ruled out
since (i) low-mass pulsars could be ejected from binary systems
and thus be isolated, and (ii) low-mass pulsars might not have
been noted and investigated comprehensively in binaries (e.g.,
pulsar/white dwarf systems, pulsar/planet systems).

There could be some candidates of low-mass quark stars (Xu 2005b).
The planck-spectrum-fitted radiation radii of CCOs (Pavlov et al.
2003) and the seven DTNs (Mereghetti et al. 2002, Haberl 2005) are
only a few kilometers. An intuition way to understand these
observations could be that these compact stars are actually
low-mass quark stars with small radii.
According to its peculiar timing behavior, it is proposed that the
radio-quiet CCO, 1E 1207.4-5209, could be a low-mass quark star
with polar surface magnetic field $\sim 6\times 10^{10}$ G and a
few kilometers in radius.
Because of color-confinement, quark stars could spin much faster
than normal neutron stars, especially for quark stars with very
small radii. It is suggested that the fastest rotating pulsar (Xu
et al. 2001), and maybe part of other millisecond radio pulsars,
could be low-mass quark stars.
The conventional way to estimate the polar magnetic field should
be revised if part of both rotation- and accretion-driven pulsars
are of low masses.

It is worth noting that the gravitational wave from pulsars should
be {\em mass}-dependent. Based on the data from the second LIGO
science run (Abbott et al. 2005) for 28 radio pulsars (3 normal
pulsars and 25 millisecond pulsars), the upper limits of
ellipticities of these pulsars are obtained (order of $10^{-5}$
for millisecond pulsars) with an assumption that pulsars have
typical masses of $\sim M_\odot$. Owen (2005) concluded that the
maximum solid strange quark star ellipticities are comparable to
the upper limits obtained by LIGO, while maximum ellipticities of
hybrid stars will be detectable by LIGO at initial design
sensitivity.
However, if pulsars are solid quark stars with approximate
Maclaurin spheroids\footnote{%
This approximation is good for the only well-studied solid
celestial body, the Earth.
}, %
one can only obtain the upper limits of $R\cdot \theta^{1/5}$
($\theta$: wobble angle) for these millisecond pulsars (Xu 2005c).
In case of $\theta\sim $ a few degrees, the 25 millisecond pulsars
could be quark stars with only a few kilometers in radius, to be
comparable to that of X-ray thermal radii detected in CCOs and
DTNs.

Searching sub-millisecond pulsars could be an expected way to
provide clear evidence for (low-mass) quark stars.
Normal neutron stars can {\rm not} spin with periods being less
than $\sim 0.5M_1^{1/2}R_6^{-3/2}$ ms ($R_6=R/10^6$ cm), but
low-mass bare strange stars can, even less than 0.1 ms.
We need thus a much short sampling time, and would deal with then
a huge amount of data in order to find a sub-millisecond pulsars.
Due to its large receiving area and wide scanning sky, the future
radio telescope, FAST (five hundred meter aperture spherical
telescope), to be built in Yunnan, China could have this chance
(to find {\em fast} pulsars via {\em FAST} \^\_\^~).

\section{Magnetospheric activity of quark stars}

As speculated in \S2, we propose pulsars to be almost homogenously
magnetized rotators, with non-zero inclination angles between
spinning and magnetic axes (Fig. 1).
\begin{figure}
  \centering
    \includegraphics[width=6cm]{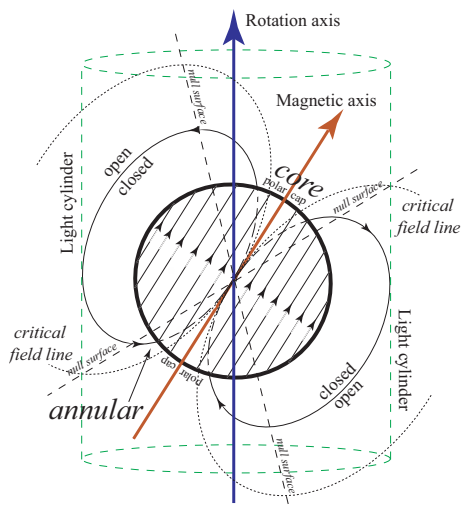}
    \caption{%
A schematic illustration of an isolated (i.e., without accretion)
quark star and its magnetospheric geometry. The quark star is
suggested to be homogeneously magnetized due to the speculation of
magnetic symmetry broken spontaneously in solid quark matter as
the star cools to a temperature to be lower than a critical Curie
temperature.
    }%
\end{figure}
For electromagnetism-torqued pulsars without accretion
environments, the closed region in their magnetospheres are
limited by the light cylinders with radius of $cP/(2\pi)$. In case
of accretion (in a binary or with a debris disk), the closed field
lines is limited by a magnetospheric radius, $r_{\rm m}$, rather
than the light cylinder radius, $r_{\rm lc}=c/\Omega$.
The charge density near a pulsar in its magnetosphere is
$\rho_{\rm gj}\simeq -{\bf \Omega\cdot B}/(2\pi c)$, which is zero
at the null surface.
Lines with $\rho_{\rm gj}=0$ at light cylinder are called {\em
critical field lines}.
Relativistic pair plasma may flow out along open filed lines,
which could result in lighthouses of multi-waveband (from radio to
$\gamma$-ray) emissions.

It is believed that primary pairs are produced and accelerated in
regions (gaps) with a strong electric field along the magnetic
line ($E_\parallel$) while more secondary pairs (with multiplicity
$\sim 10^2-10^4$) are created outside the gaps ($E_\parallel\neq
0$). Numerous models have been suggested concerning gap
acceleration, which are classified into two groups: vacuum inner
and outer gaps (e.g., Ruderman \& Sutherland 1975, Cheng, Ho, \&
Ruderman 1986), and free-flow gaps (e.g., Arons \& Scharlemenn
1979; Harding \& Muslimov 1998).
The former depends on strong binding energy of charged particles
on the pulsar surface, but the later on negligible one.
Since different radiative characters could be expected in those
two groups of models, we may probe into the nature of particles on
pulsar surface, and thus of pulsar interior, through emission
features.

Both positively and negatively charged particles on surfaces of
normal neutron stars can not be enough bound for vacuum gaps, but
that of bare quark stars can.
We thus expect vacuum inner gaps work for quark star's
magnetospheric activities (see next about the outer gap).
Sparking process is a direct consequence of vacuum gaps, which
results naturally inhomogeneous plasma ejecta in open field-lines.
The electrodynamic description to calculate the acceleration of
pairs would then not easily be formulated mathematically.
Observations in favor with the above scenario of polar-cap
sparking include short time-scale events (microstructure pulses,
nanosecond giant radio bursts), drifting subpulses, and no
intrinsic rotation measure
($ {\rm RM} \simeq 0.81\int_{\rm PSR}^\oplus n_e{\bf B}\cdot d{\rm
s}$)
detected in pulsar magnetospheres.

The region of open-field lines can be divided into core and
annular ones. The boundary of the former is all of the critical
field lines, while the later is between the surfaces of critical
and last-open field lines.
Sign-changed plasma would flow out from these two regions in order
to close the global electric circuit of magnetosphere.
Two inner vacuum gaps (inner core and annular gaps) may occur,
which could explain the pulse profiles of  multi-band emission
(Qiao 2004a) and the bi-drifting Phenomenon (Qiao 2004b).
There is a competition between the inner annuler gap and the outer
gap. If pair production favors near polar caps due to high opacity
of photons and strong $B$ and $E_\parallel$, dense pair plasma
ejected may quench the outer gap.

A pulsars could be monopole-charged with an electricity, $Q$,
depending on its radius $r$, polar magnetic field $B$, and spin
(Xu et al. 2006).
The sign-changed plasma could be separated by critical field lines
if $Q\sim 10^{-3} (r/10^6{\rm cm})^3(B/10^{12}{\rm G})/P^2$
Coulomb.
However, if a pulsar has not such charge, the core and annular
regions would not be separated by the critical field lines. This
could consequently affect the radiative locations of radio as well
as higher energy emission, which might be checked by the
multi-wavelength observations of pulse profiles.

\section{Conclusions}

Quark matter and quark stars are reviewed at an astrophysical
viewpoint. In addition to test Einstein's general relativity in
strong field, pulsar-like stars are also useful to test and to
improve the fundamental strong interaction. Some characteristics
of quark stars are proposed, which may lead finally to a
successful identification of them. The physics relevant to the
elementary chromatic interaction would certainly be improved if
pulsar-like stars are really quark stars. Based on variety of
astrophysical observations, a solid state of quark matter is
suggested.


\vspace{5mm}%
\noindent%
{\em Acknowledgments}:
The author thanks helpful discussion with the members in the
pulsar group of Peking University. This work is supported by NSFC
(10273001) and the Key Grant Project of Chinese Ministry of
Education (305001).


\end{document}